\documentclass[fleqn,10pt]{wlscirep}
\title{Magnetic resonance as a local probe for kagom\'{e} magnetism in Barlowite Cu$_4$(OH)$_6$FBr}

\author[1,*]{K. M. Ranjith}
\author[2]{C. Klein}
\author[3]{A. A. Tsirlin}
\author[1]{H. Rosner}
\author[2]{C. Krellner}
\author[1]{M. Baenitz}
\affil[1]{Max- Plank- Institute for Chemical Physics of Solids, 01187 Dresden, Germany}
\affil[2]{Physikalisches Institut, Goethe-Universitat Frankfurt, 60438 Frankfurt, Germany}
\affil[3]{Experimental Physics VI, Center for Electronic Correlations and Magnetism, Institute of Physics, University of Augsburg, 86135 Augsburg, Germany}
\affil[*]{ranjith.kumar@cpfs.mpg.de}

\begin{abstract}
Temperature- and field- dependent $^1$H-, $^{19}$F-, and $^{79,81}$Br- NMR measurements together with zero - field $^{79,81}$Br-NQR measurements on polycrystalline samples of barlowite, Cu$_4$(OH)$_6$FBr are conducted to study the magnetism and possible structural distortions on a microscopic level. The temperature dependence of the $^{79,81}$Br- NMR spin-lattice relaxation rates 1/$T_1$ indicate a phase transition  at $T_{\rm N}\simeq$15 K which is of magnetic origin,  but with an unusually weak slowing down of fluctuations below $T_{\rm N}$.  Moreover, 1/$T_1T$ scales linear with the bulk susceptibility which indicates persisting spin fluctuations down to 2 K. Quadupolare resonance (NQR) studies reveal a pair of zero-field NQR- lines associated with the two isotopes of Br with the nuclear spins of $I$ = 3/2. Quadrupole coupling constants of $\nu_Q\simeq$ 28.5~MHz and 24.7~MHz for $^{79}$Br- and $^{81}$Br- nuclei are determined from Br-NMR and the asymmetry parameter of the electric field gradient was estimated to $\eta \simeq 0.2$. The Br-NQR lines are consistent with our findings from Br-NMR and they are relatively broad,  even above $T_{\rm N}$. This broadening and the relative large $\eta $ value suggests a symmetry reduction at the Br- site reflecting the presence of a local distortion in the lattice. Our density-functional calculations show that the displacements of Cu2 atoms located between the kagome planes do not account for this relatively large $\eta$. On the other hand, full structural relaxation, including the deformation of kagome planes, leads to a better agreement with the experiment.
\end{abstract}
\begin{document}

\flushbottom
\maketitle

\thispagestyle{empty}

\section*{Introduction}

Magnetic frustration, the competing interactions between the spins enhances the quantum fluctuations and leads to exotic ground states and novel phases. One of the important manifestation is the formation of the quantum spin liquid (QSL) ground state,\cite{Balents2010,Savary2017,Zhou2017,Powell2011} a highly entangled spin state without any symmetry breaking even at zero temperature. The important reported examples include the two-dimensional kagome lattice quantum magnet Herbertsmithite ZnCu$_3$(OH)$_6$Cl$_2$,\cite{Helton2007,Han2012,Norman2016} the triangular lattice organic magnets $\kappa$–(BEDT-TTF$_2$Cu$_2$(CN)$_3$ and  EtMe$_3$Sb[Pd(dmit)$_2$]$_2$,\cite{Yamashita2008,Yamashita2009} and three-dimensional hyperkagome iridate Na$_4$Ir$_3$O$_8$.\cite{Singh2013}  The quest for more experimental QSL candidates has triggered the intense research in two dimensional kagome lattice systems with spin-$\frac{1}{2}$ connectivity.

\begin{figure}[ht]
{\centering {\includegraphics[width=1\textwidth]{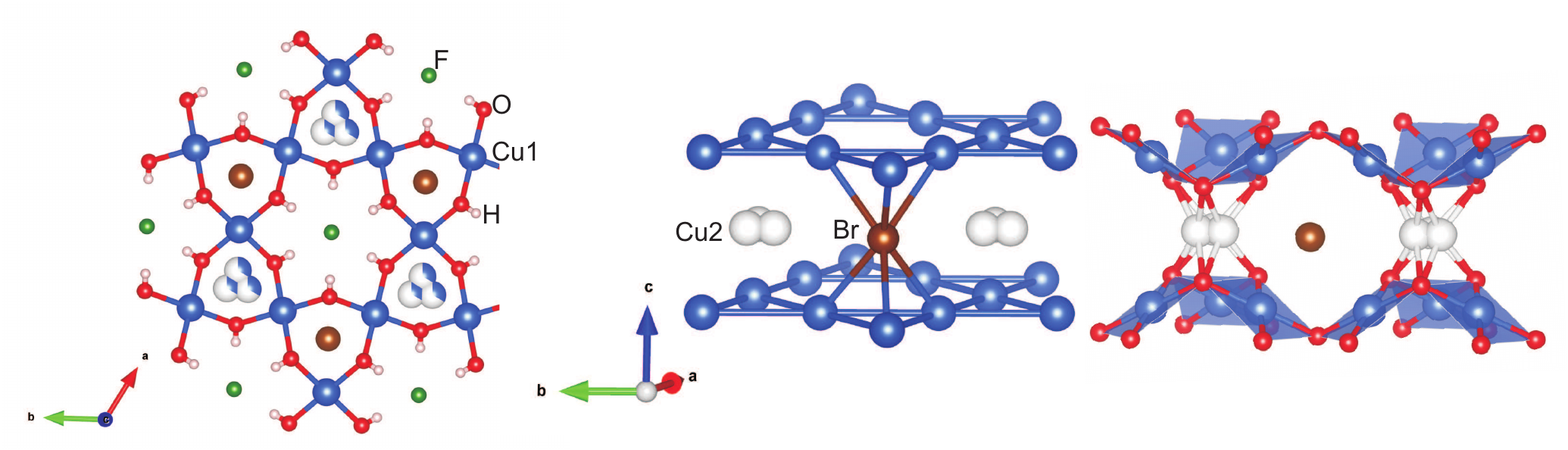}}\par} \caption{(Color online) The crystal structure of barlowite explaining the local environment of Br together with the intralayer Cu1- position and the distorted (three equivalent sites possible) interlayer Cu2- position.} \label{structure}
\end{figure}

Barlowite, Cu$_4$(OH)$_6$FBr is a structural analog of Herbertsmithite which recently has attracted much attention as a geometrically perfect spin-$\frac{1}{2}$ kagome antiferromagnet due to an absent antisite disorder (Zn – Cu).\cite{Han2014,Jeschke2015,Liu2015,Gati2015,Han2016,Guterding2016} Barlowite crystalizes in a hexagonal structure (space group $P6_3/mmc$) with lattice parameters $a\simeq6.6786~\AA$ and $c\simeq 9.2744~\AA$. As shown in Fig.~\ref{structure}, crystallographically equivalent three Cu(1)$^{2+}$- ions in the chemical formula form the kagome layer in the $ab$-plane, whereas the fourth Cu(2)$^{2+}$- ion in the chemical formula occupies the inter-layer sites. The interlayer site is quite large and the fourth Cu(2)$^{2+}$- ion occupies one out of the three possible equivalent positions which might originate some local distortion of the lattice and a symmetry breaking.

The lattice structure of barlowite is close to that of Herbertsmithite, but has a different stacking pattern of kagome layers because of the different bonding environment of halogen ions. In Herbertsmithite, Cl$^{-}$- ions form both covalent and hydrogen bonds and this lead to a horizontal staggering of kagome layers. In the case of barlowite, Br$^{-}$ forms six covalent bonds with Cu$^{2+}$- ions whereas F$^{-}$- ions form six hydrogen bonds and features a perfect alignment of kagome layers on top of each other. The magnetic exchange between the intralayer (Cu1) and the interlayer (Cu2) Cu$^{2+}$- ions in barlowite is rather complex.\cite{Jeschke2015} The low temperature magnetism and in particular the magnetic order is determined by the presence of the interlayer Cu2 ions and their ferromagnetic coupling to the intralayer Cu1 ions. In the antiferromagnetic kagome layer the Cu1 moments are highly frustrated with a rather high coupling constant of about 170 K. Canted antiferromagnetic order at $T_{\rm N}\approx$ 15~K is revealed from bulk probes at small magnetic fields \cite{Han2014} wheras at higher fields a significant ferromagnetic contribution is observed in the bulk magnetization.  Susceptibility measurements on polycrystalline material provides a large antiferromagnetic Curie-Weiss temperature of $\theta_{\rm CW}$ = -136$\pm$10~K with a small hysteresis in the magnetization curve (see supplemental material). The possibility of a QSL ground state (of Cu1 Kagome moments) coexisting with ordered Cu2 interlayer moments is discussed in the literatures.\cite{Han2014, Han2016} Recent studies for barlowite on the Zn substitution on the Cu2 site support this idea. Here magnetic order is suppressed and the fully substituted Zn barlowite Cu$_3$Zn(OH)$_6$FBr is considered as a QSL system \cite{Guterding2016,Feng2017,Yuan2018} very similar to the fully Zn substituted Herbertsmithite Cu$_3$Zn(OH)$_6$Cl$_2$ .

The nuclear magnetic resonance (NMR) is a microscopic probe for static and dynamic magnetism and was successfully applied on Herbertsmithite and its polymorphs \cite{mendels2016,bert2009,lacroix2013} but so far due to the lack of sizable  amounts of pure polycristalline material no NMR study was conducted on barlowite. Whereas the moment transfer from the magnetic ion originates the NMR shift $K$ which provides information about the intrinsic static spin susceptibility $\chi (q=0, \omega = 0)$ and  the spin-lattice relaxation rate 1/$T_1$ supplies the information about excitations in particular about the dynamic complex susceptibility $\chi'(q,\omega)$. In this paper, we use magnetic resonance as a local probe for magnetism and simultaneously for the local structure around the NMR active nuclei. In contrast to NMR studies on herbertsmithite, barlowite contains a novel NQR nuclei (Br) which hosts a large quadrupole moment and therefore allows for rather sensitive NQR studies as a measure of lattice distortions and to evidence the claim of the perfect kagome motif.  In general, the nuclear spin resonance spectra of nuclei with $I>\frac{1}{2}$ contains additional information (beside the magnetic interaction mentioned before) due to the presence of the quadrupole interaction. Here the $I>\frac{1}{2}$ NMR active nuclei has a quadrupole moment which is exposed to the electric field of the surrounding ions (frequently modelled by the point charge model). Therefore the quadrupole interaction depend in first place on the electric field gradient at the nucleus site (charge distribution around the nucleus) and the quadrupole moment of the nucleus itself. As for NMR the main interaction is the magnetic one (Zeeman interaction) and the quadrupole interaction is just a perturbation the NQR technique (at zero field) is more favorable to probe the quadrupole interaction solely. Two $I=\frac{3}{2}$- isotopes with 100\% natural abundance in total and relatively large quadrupole moments on both nuclei identifies Br- NMR and NQR as a powerful technique which is frequently applied on quantum spin systems and other forms of matter.\cite {Pregelj2013,Comment2010,Baek2012,Rebecca2009}  Our Br magnetic resonance study presented here aims for the consistent determination of the quadrupole coupling constant $\nu_{\rm Q}$ and the asymmetry parameter $\eta$ from NMR and NQR.

\section*{Results and Discussion}

\begin{figure}[ht]
{\centering {\includegraphics[width=0.65\textwidth]{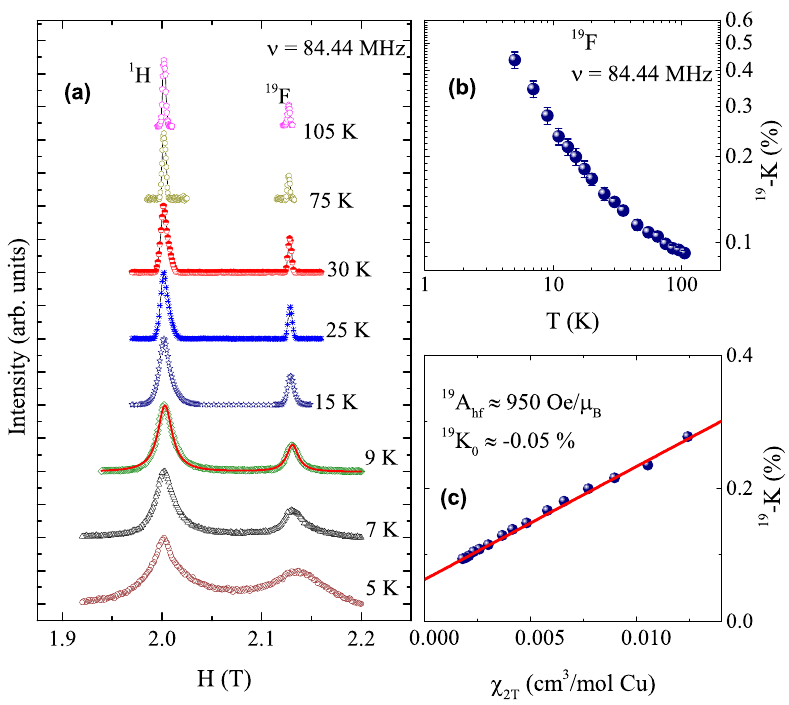}}\par} \caption{(Color online) Field-sweep $^1$H- and $^{19}$F- spectra of barlowite measured at $\nu$ = 84.44~MHz at various temperatures (a).$^{19}$F-NMR shift as a function of temperature. Red solid line is the two Lorentzian fit (b).  $^{19}$F-NMR shift versus $\chi_{\rm bulk}$ plot with T as implicit parameter (solid line indicates the linear behaviour) (c). } \label{spk}
\end{figure}

Figure~\ref{spk} shows the field sweep $^1$H- and $^{19}$F-NMR spectra measured at a fixed frequency of 84.44~MHz for different temperatures. No shift in the central peak position was observed for the $^1$H- line, whereas a small shift was found for the $^{19}$F- line. Both NMR lines show a significant broadening below 15~K which is consistent with the magnetic order at 15~K (see supplementary materials Fig. S1).\cite{Han2014}  The NMR Knight shift $K(T)= [H_{\rm ref}-H^*(T)]/H^*(T)$ was determined by measuring the resonance field  of the sample, $H^*(T)$ as a function of the temperature with respect to the nonmagnetic reference sample (resonance field $H_{\rm ref}$). The temperature-dependent $K(T)$ of $^{19}$F line is presented in the Fig.~\ref{spk}(b). It shows a Curie-Weiss behaviour similar to the $\chi(T)$ data (see supplemental material). Generally, $K(T)$ is directly proportional to the spin susceptibility ($\chi_{\rm spin}$) and can be written as $K(T)=K_0+\frac{A_{\rm hf}}{N_A}\chi_{\rm spin}$, where $K_0$ is the temperature independent orbital (chemical) shift, $N_{\rm A}$ is the Avogadro number, and $A_{\rm hf}$ is the hyperfine coupling constant between the $^{19}$F  nucleus and the Cu$^{2+}$ magnetic moments. The conventional scheme for calculating $A_{\rm hf}$ is to take the slope of the $K$ vs. $\chi$ plot with $T$ as an implicit parameter. As shown in Fig.~\ref{spk}(c), the $K$ vs. $\chi$ plot is fitted very well by a linear behaviour over a wide temperature range yielding $A_{\rm hf}\approx$ 950~Oe/$\mu_{\rm B}$ and $K_0\approx$ -0.05~\%.

\begin{figure}[ht]
{\centering {\includegraphics[width=0.65\textwidth]{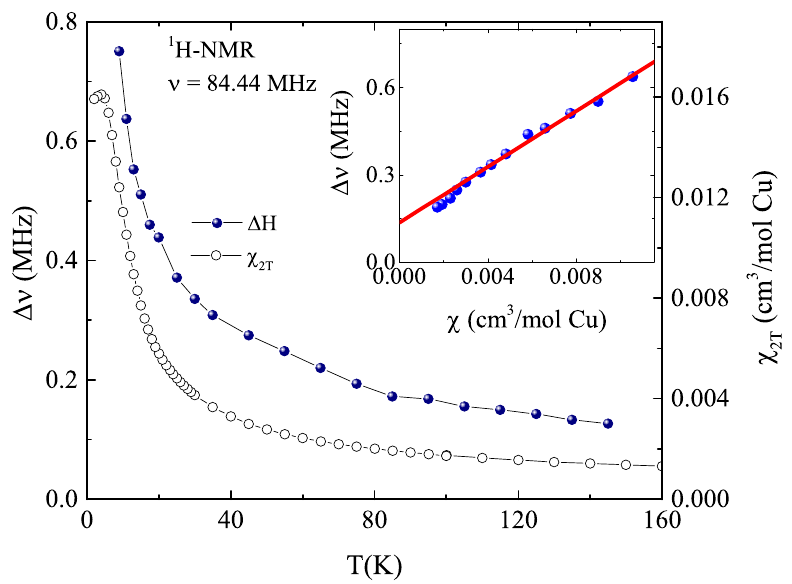}}\par} \caption{(Color online) Temperature dependence of full width at half maximum ($\Delta \nu$) of the $^1$H NMR spectra measured at 84.44~MHz (left axis) and magnetic susceptibility ($\chi$) measured at $H=$ 2~T (right axis). The inset shows the line width $\Delta \nu$ vs. $\chi$ with temperature as an implicit parameter and the solid line is the linear fit.} \label{fwhm}
\end{figure}

The $^1$H- and $^{19}$F- NMR lines were found to broaden monotonously upon lowering the temperature. The $^1$H- NMR line width $\Delta \nu = \frac{\gamma}{2\pi}\Delta H$ (defined as the full width half maximum, FWHM) as a function of temperature is shown in Fig.~\ref{fwhm}. $\Delta \nu$ is increasing systematically with decreasing temperature and tracks the magnetic susceptibility. For a comparison, in the same Fig.~\ref{fwhm}, $\chi(T)$ measured at 2~T is also plotted in the right $y$-axis. The NMR spectral width has two major contributions, (i) the dipolar interaction between the nuclei and (ii) the dipolar interaction of the $^1$H- nuclei with the neighboring Cu$^{2+}$ magnetic ions and can be expressed as,
$\Delta\nu = \Delta\nu_0+\Delta \nu_{\rm aniso}$. The first term is a temperature and field independent contribution from the nuclear-nuclear dipolar interaction, whereas the second term is the powder average over the anisotropic dipolar interaction between the$^1$H- nuclei and magnetic Cu$^{2+}$- ions which is $T$- dependent. In an external magnetic field,  this anisotropic coupling gives rise to inhomogeneous line broadening which is proportional to the bulk
magnetic susceptibility $\chi(T)$. So one can write as,\cite{Procissi2004,Suh2003,Ranjith2015,Belesi2007} $$\frac{(\Delta \nu)_{\rm aniso}}{\nu_{\rm L}} = A_{\rm z}\chi \approx \frac{<\mu>}{r^3H}$$

where $\nu_L = \frac{\gamma_{\rm N}}{2\pi}H$ denotes the $^1$H Larmor frequency for an applied external magnetic field $H$, $A_{\rm z}$ is the average dipolar coupling constant between the $^1$H nucleus and magnetic Cu$^{2+}$ ions, and $r$ is the average distance between them. Finally, we can rewrite \begin{equation}
\Delta \nu = \Delta \nu_0 + \frac{\gamma_{\rm N}}{2\pi}A_{\rm z} H\chi.
\label{width}
\end{equation}

 In the inset of Fig.~\ref{fwhm} we plotted $\Delta \nu$ versus $\chi$ with temperature as an implicit parameter and it follows a linear behaviour. This linear  $\Delta \nu$ vs. $\chi$ is fitted by Eq.~\ref{width} and the obtained value of dipolar coupling constant $A_{\rm z} \sim$10$^{22}$\, cm$^{-3}$ corresponds to the field generated by the Cu$^{2+}$- moments at an average distance of 3~$\AA$. This suggests that the interaction of the $^1$H- nucleus with Cu$^{2+}$ moments is largely of dipolar origin.

\begin{figure}[ht]
{\centering {\includegraphics[width=0.7\textwidth]{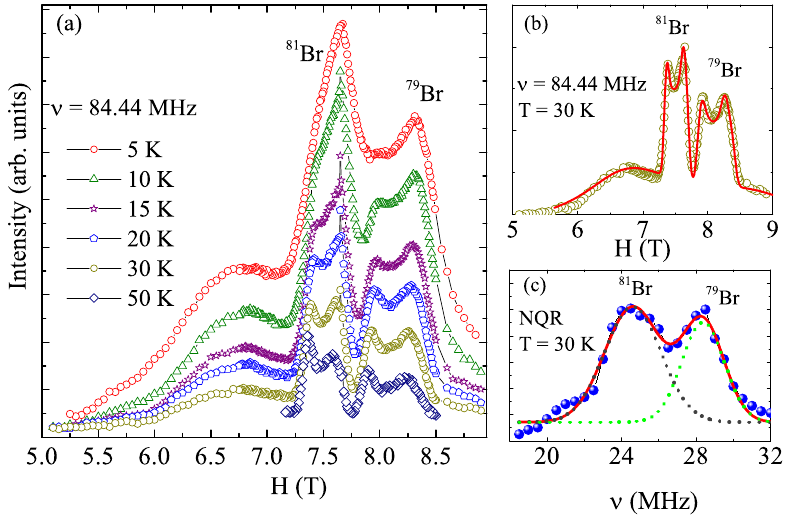}}\par} \caption{(Color online) (a) $^{79}$Br- and $^{81}$Br- spectra of barlowite measured at $\nu$ = 84.44~MHz for different temperatures.
 (b) $^{79}$Br- and $^{81}$Br- spectrum at $T=$ 30~K together with a simulation (solid line, parameters see text). (c) Corresponding NQR spectrum at $T=$ 30~K simulated with the same parameter as used for the NMR spectrum (solid lines and dashed lines).} \label{Br_spk}
\end{figure}

Figure~\ref{Br_spk}(a) shows the field sweep Br- NMR spectra measured at 84.44~MHz at different temperatures. Multiple lines are found corresponding to both $^{79}$Br- and $^{81}$Br- isotopes exposed to first- and second- order quadrupole interaction and with powder averaged intensity. At high temperatures,  Br- sites yield a complex NMR spectra consist of splitted central lines with broad satellites. The central line arises from the (+$\frac{1}{2}\rightarrow -\frac{1}{2}$) transition split into two peaks due to the combined effect of magnetic anisotropy and second order quadrupolar interaction. No significant shift was observed with temperature, which reflects a weak transferred hyperfine coupling of the $^{79,81}$Br nuclei with the Cu$^{2+}$ spins. The spectra starts to broaden with decrease in temperature. Both $^{79}$Br- and $^{81}$Br- are quadrupole nuclei with nuclear spin $I=\frac{3}{2}$ and gyromagnetic ratio $\gamma/2\pi=$ 10.666 MHz/T and 11.498~MHz/T, respectively. For a quadrupole nuclei, the nuclear spin Hamiltonian can be expressed as a summation of the Zeeman interaction and the quadrupole interaction as $H= H_{\rm Z} + H_{\rm Q}$, where $H_{\rm Z} = -\gamma \hbar I H_{\rm eff}$ represents the zeeman term and  $H_{\rm Q} = \frac{h\nu_{\rm Q}}{6}[3I^2_Z-I(I+1)+\frac{1}{2}\eta(I_x^2-I_y^2)]$. $H_{\rm eff}$ is the effective field which is the sum of external field $H$ and the hyperfine field $H_{\rm hy}$ at the Br site, $\nu_Q = 3eQV_{zz}/2hI(2I-1)$ is the quadrupole coupling constant, $V_{zz}$ is the largest component of electric field gradient (EFG) tensor, and $eQ$ is the nuclear quadrupole moment. The EFG tensor is generally defined as $V_{zz}|\geq|V_{yy}|\geq|V_{xx}|$ with $\eta=(V_{xx}-V_{yy})/V_{zz}$ as the asymmetry parameter.\cite{Slichter1990,Abragam1996}  When the quadrupole term in the Hamiltonian is weak compared to the Zeeman term, the quadrupolar term modifies the NMR spectrum and can be treated as a first order perturbation. In this case, the position of the central line ($\frac{1}{2}\leftrightarrow -\frac{1}{2}$ transition) is determined by the applied magnetic field whereas the distance between the satellite lines (corresponding to the $-\frac{3}{2}\leftrightarrow -\frac{1}{2}$ and $\frac{3}{2}\leftrightarrow \frac{1}{2}$ transitions)  depend on the angle $\theta$ between the magnetic field direction and the direction of the maximum of the EFG component $V_{zz}$. When quadrupole effects are considered to second order (and for axial symmetry), the central line position (in the absence of anisotropy) also depends on $\theta$ and is given by
\begin{equation}\label{quad}
  \nu_{(\pm \frac{1}{2})} = \nu_0 + \frac{\nu_Q^2}{32\nu_0}\left[I(I+1)-\frac{3}{4}\right](1-cos^2\theta)(9cos^2\theta-1),
\end{equation}

where $\nu_0$ the Larmor frequency. For a randomly aligned polycrystalline sample the central line shape is the powder average, resulting in two peaks corresponding to $\theta\simeq 41.8 ^{\circ} $  and $\theta\simeq 90^{\circ}$.\cite{Grafe2008}

At 30~K, the spectra was fitted reasonably well, taking into account the second-order quadrupolar perturbation contribution. Figure~\ref{Br_spk}(b) shows the measured spectra at 30~K together with the theoretical simulation.  The parameters obtained are $K_{\rm iso}\simeq -0.73~\%$, $K_{\rm ax}\simeq 0.185~\%$, $\eta \simeq 0.2$, and $\nu_Q\simeq$ 28.5~MHz for $^{79}$Br line and $K_{\rm iso}\simeq -0.74~\%$, $K_{\rm ax}\simeq 0.2~\%$, $\eta \simeq 0.2$, and $\nu_Q\simeq$ 24.7~MHz for $^{81}$Br line. As both isotopes are exposed to the same electric field gradient, the ratio $^{79}\nu_Q/^{81}\nu_Q$ gives the ratio of their qudrupole moments.

Both the isotopes of bromine are quadrupole nuclei and have large quadrupole moment ($^{79}Q$ = 0.33$\times10^{-28}$~m$^2$ and $^{81}Q$ = 0.28$\times10^{-28}$~m$^2$) and almost same natural abundance. For spin-$\frac{3}{2}$ nuclei, which have two doubly degenerate energy levels, $$E_{\pm\frac{3}{2}} = \frac{h\nu_Q}{2}(1+\frac{\eta^2}{3})^\frac{1}{2}, E_{\pm\frac{1}{2}} = -\frac{h\nu_Q}{2}(1+\frac{\eta^2}{3})^\frac{1}{2}$$  and the only one pure quadrupole resonance frequency can be obtained as $\nu_{\rm NQR} = \nu_Q(1+\eta^2/3)^\frac{1}{2}$.\cite{Das1958}
The NQR spectrum measured at 30~K is shown in Fig.~\ref{Br_spk}(c). We observed a broad spectrum which is the combination of two lines at $\nu_{\rm NQR}$ =  28.3~MHz and 24.5~MHz corresponding to the $^{79}$Br and $^{81}$Br quadrupole resonance, respectively.   The ratio of the resonance frequencies and the intensities of the lines correlate well with the ratio of the quadrupole moments and the natural abundance for $^{79}$Br and $^{81}$Br isotopes, respectively. The resonance frequencies are also consistent with the parameter obtained from the NMR spectral simulation at 30~K.

Br atoms occupy the $2c$ position with the $\bar 6m2$ symmetry in the hexagonal structure of barlowite, so their EFG asymmetry should be zero. The non-zero $\eta$ value is indicative of deviations from the hexagonal symmetry caused by the displacement of Cu2 atoms away from the three-fold axis. The influence of this displacement on the electric field gradient was explored by density-functional (DFT) calculations that arrived at $\eta=0.08$ for Br. Local displacements of Cu2 may also affect the positions of other atoms, resulting in deviations from the average hexagonal crystal structure. To gain further insight into these structural changes, we performed full relaxation and optimized all atomic coordinates until energy minimum was reached. This led to two possible configurations (see Methods) with slightly different $\eta$'s of 0.12 and 0.18, both approaching the experimental $\eta\simeq 0.2$. We thus show that the displacements of Cu2 do not account for the full experimental EFG asymmetry. The displacements of others atoms contribute as well. The local structure of barlowite may not be well described by the average hexagonal model, a conclusion that goes in line with the very recent diffraction studies reporting orthorhombic symmetry of the crystal structure.\cite{Pasco2018,Feng2018}

\begin{figure}[ht]
{\centering {\includegraphics[width=0.7\textwidth]{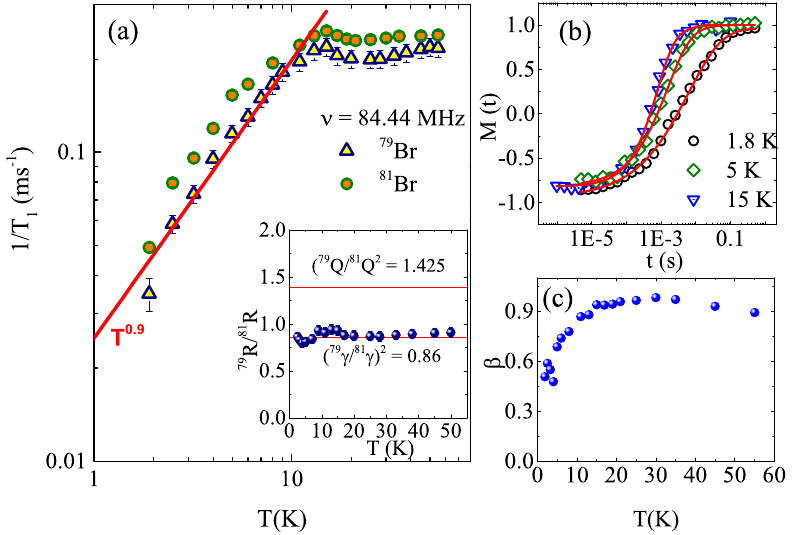}}\par} \caption{(Color online) (a) The temperature dependent spin-lattice relaxation rate (1/$T_1$) for $^{79}$Br and $^{81}$Br in barlowite. The red solid line corresponds $T^{0.9}$.  The inset shows the ratio of the spin-lattice relaxation rates for $^{79}$Br and $^{81}$Br measured at same external field. Horizontal solid lines indicate the corresponding ratio for pure quadrupolar and pure magnetic relaxation mechanism. (b) Recovery of the longitudinal magnetization as a function of delay time along with the theoretical fit (solid line) for three different temperatures. (c) Temperature dependence of the stretch exponent $\beta$ (see text).  } \label{T1}
\end{figure}

Insights into the spin dynamics can be obtained by measuring the spin lattice relaxation rates ($1/T_1$) as a function of temperature. $^{79}$Br and $^{81}$Br nuclear spin-lattice relaxation time $T_1$ were measured using a conventional inversion recovery pulse sequence at 84.44 MHz (at 7.4~T for $^{81}$Br and 8.2~T for $^{79}$Br). Values of $T_1$ were obtained from fits to an appropriate relaxation function. The recovery of the longitudinal magnetization $M(t)$ at a time delay $t$ after the inversion pulse could be fitted consistently with a single $T_1$ function;
$(1-\frac{M(t)}{M(\infty)}) = 0.1~\rm{exp}[-(t/T_1)^\beta]+0.9~\rm{exp}[-(6t/T_1)^\beta]$, where $\beta$ is the
stretch exponent. The temperature dependence of $1/T_1$ is shown in Fig~\ref{T1}(a). At high temperatures ($T\geq 20~K$), 1/$T_1$ is almost temperature-independent (in the paramagnetic regime) for both isotopes, which is often observed in a system with localized moments when the temperature is higher than the exchange energy between the spins.\cite{Moriya1956} With decreasing temperatures, 1/$T_1$ shows an anomaly at $T_{\rm N}\simeq$ 15\,K  associated with some sort of phase transition consistent with the thermodynamic measurements.\cite{Han2014} Below $T_{\rm N}$, 1/$T_1$ decreases smoothly towards zero as a result of the disappearance of the critical fluctuations in the ordered state. The overall behaviour of temperature dependence of $1/T_1$ is similar for both the isotopes. In order to understand the relaxation mechanism, we have measured $1/T_1$ for both isotopes at the same applied magnetic field. In the case of a pure magnetic relaxation mechanism, the ratio of the relaxation rates should be equal to the ratio of the squares of corresponding gyromagnetic ratio, $(^{79}\gamma/^{81}\gamma)^2\approx 0.85$. Whereas, for a pure quadrupole relaxation mechanism, the ratio of relaxation rates will be equal to the ratio of the squares of their quadrupole moments $(^{79}Q/^{81}Q)^2\approx 1.425$. In the inset of  Fig~\ref{T1}(a), we have plotted the ratio of relaxation rate of both isotopes, $^{79}R/^{81}R$ as a function of temperature, where $R$ denotes the spin lattice relaxation rate $1/T_1$. It is observed that the ratio of the relaxation rates are very close to the value $(^{79}\gamma/^{81}\gamma)^2\approx 0.85$ indicating the magnetic nature of the relaxation mechanism. Nonetheless it has to be mentioned that the behaviour of 1/$T_1$ in the ordered state is rather unconventional for a bulk ordered antiferromagnet \cite{Moriya1985} and the power law obtained ($T^{0.9}$) has a rather small exponent. A reduced power law in 1/$T_1$ was even found in Cl NMR on Herbertsmithite which does not show any long range order at all.\cite{Imai2008}  Br NMR in barlowite evidences a rather weak slowing down of critical fluctuations towards low temperatures. For Barlowite this might indicates that only a fraction of Cu magnetic moments undergoes long-range magnetic order (f.i. interlayer Cu2 moments), or the ordered moment is rather small, and significant spin fluctuations persist in the ordered state.  This is also supported by the magnetization recovery (Fig. 5b). Very recent neutron diffraction studies on the barlowite also support this scenario. Here, strong residue spin fluctuations, in the form of intralayer spin singlet formation due to the structure distortion (both displacements of Cu2 and deformation of kagome planes), are reported below magnetic transition at 15~K.\cite{Feng2018} For systems with unique and homogeneous relaxation rate, the value of the stretch exponent will be $\beta=1$. In contrast, a reduced value of $\beta (<1)$ indicates a wide distribution of relaxation rates. As shown in Fig.~\ref{T1}(c), the value of $\beta$ is close to 1 at higher temperatures, but it decreases with decreasing temperature below 20~K. This behaviour is frequently found for frustrated low dimensional quantum magnets where magnetic order competes with persistent liquid like magnetic excitations.\cite{Itou2010,Shiroka2011,Bosio2017,Shockley2015}

 \begin{figure}[ht]
{\centering {\includegraphics[width=0.7\textwidth]{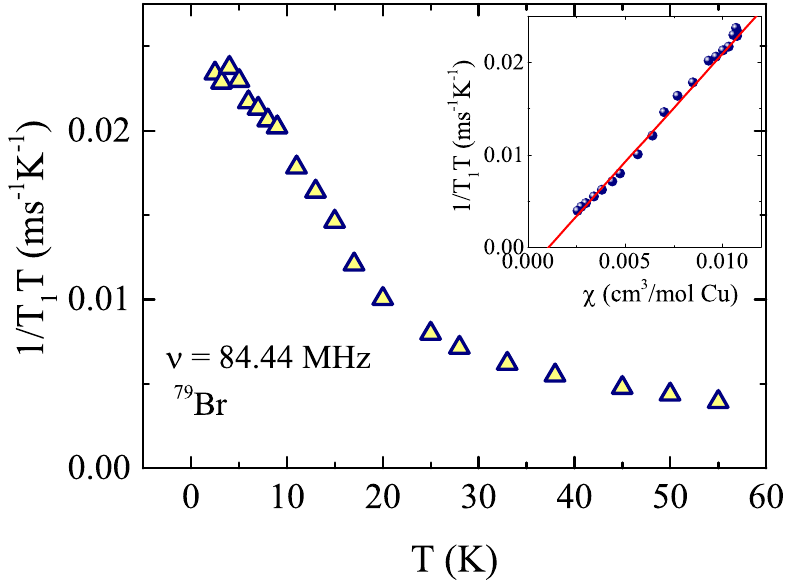}}\par} \caption{(Color online) The temperature dependence of 1/$T_1T$ for the $^{79}Br$ line. The linear relation between 1/$T_1T$ and $\chi$ with temperature as an implicit parameter is shown in the inset.} \label{TT1}
\end{figure}

As shown in Fig.~\ref{TT1}, the temperature dependence of 1/$T_1T$ gradually increases with decreasing temperature and shows a steep up-turn below around 20~K. This up-turn in 1/$T_1T$ below 20~K indicates the typical behaviour near $T_{\rm N}$  due to the critical Cu$^{2+}$ moment fluctuations. Frequently a scaling with the bulk susceptibility is observed.\cite{Khuntia2017}  The spin-lattice relaxation rate probes the $q$- averaged low energy spin excitations and can be expressed in terms of  imaginary part of the dynamic spin susceptibility $\chi^{''}_{M}(\vec{q},\omega _{0})$ as

\begin{equation}
\frac{1}{T_{1}T} = \frac{2\gamma_{N}^{2}k_{B}}{N_{\rm A}^{2}}
\sum\limits_{\vec{q}}\mid A(\vec{q})\mid
^{2}\frac{\chi^{''}(\vec{q},\omega_{0})}{\omega_{0}},
\label{t1form}
\end{equation}
where the sum is over the wave vectors $\vec{q}$ within the first Brillouin zone,
$A(\vec{q})$ is the form factor of the hyperfine interactions as a
function of $\vec{q}$, and $\omega _{0}$ is the nuclear Larmor frequency .\cite{Moriya1963}
In the strict (ferromagnetic) limit of $q=0$ and $\omega_{0}=0$, the real component of $\chi_{M}^{'}(\vec{q},\omega _{0})$ corresponds to the uniform static bulk susceptibility $\chi$ and  1/$T_1T$ $\propto \chi\tau$ with $\tau$ being the on site fluctuation rate is valid.\cite{Khuntia2017} In the inset of Fig.~\ref{TT1}, we show 1/$T_1T$ against $\chi$ with temperature as an implicit parameter. The linear relation between 1/$T_1T$ and $\chi$ indicates a constant on site fluctuation rate $\tau$ over a wide range in temperature and even trough the transition. This reflects the fact that barlowite is far from being a regular antiferromagnet in particular due to persisting magnetic fluctuations down to low temperatures. It still remains an open question what kind of magnetic order occurs in barlowite below 15~K, and if all Cu moments are part of that order or if for instance only the inter layer Cu2 moments participate and the Cu1 moments in the kagome plane remain fluctuating and spin liquid like towards low temperatures. This scenario is supported by the small entropy found across the transition and the small saturation moment.\cite{Han2016}  Furthermore there are ferro and antiferromagnetic exchange couplings among the two different Cu moments accompanied by a lattice anisotropy.\cite{Jeschke2015}  The static and dynamic (this means excitations) magnetism in barlowite is far from being understood and there is a strong demand for other local probes like neutron scattering or spin Muon resonance ($\mu$SR) and for studies on single crystals. As for all herbertsmithite relatives (like barlowite or kapellasite) and planar kagome-like minerals (like volborthite or vesignieite) a combination of local probes is required to disentangle between spin freezing, long-range order and the QSL ground state (see review of P. Mendels and S. Willis\cite{lacroix2013}).

\section*{Summary}

In conclusion, we investigated the magnetism in the spin-$\frac{1}{2}$ quantum magnet barlowite Cu$_4$(OH)$_6$FBr via the microscopic NMR and NQR techniques.  $^1$H, $^{19}$F, and $^{79,81}$Br NMR measurements are performed in a wide temperature and field range together with Br NQR measurements. $^1$H and $^{19}$F NMR results reveal well defined NMR lines and hyperfine coupling parameters. A broadening of the line width of the $^1$H, $^{19}$F, and $^{79,81}$Br NMR lines indicate some form of order below 15~K. Spin lattice relaxation rate measurements performed on both Br isotopes reveal a kink which probably indicate magnetic ordering at $T_{\rm N}\simeq 15~K$ and which is solely magnetic in origin. The temperature dependence of the spin lattice relaxation rate below 15~K is rather weak ( $T^{0.9}$ power law) and evidences persisting magnetic fluctuations even in the ordered state. NMR studies on single crystals together with inelastic neutron studies would be required to probe the complex magnetic structure below 15K. Furthermore we performed a comparative NMR and NQR study on both Br isotopes.  We have observed a pair of NQR lines corresponding to the $^{79,81}$Br nuclei and in good agreement with the complex Br NMR spectra obtained.  From our $^{79,81}$Br NMR and NQR measurements, we obtain an asymmetry parameter of the electric field gradient $\eta \simeq 0.2$ and the  quadrupole coupling constants $\nu_Q\simeq$ 28.5~MHz and 24.7~MHz  for $^{79}$Br and $^{81}$Br nuclei, respectively. The broad NQR spectrum observed at 30~K and the relatively large electric field gradient indicate a symmetry reduction at the Br site which can not be explained by the displacements of Cu2 alone and indicates salient deviations of the barlowite structure from the hexagonal symmetry. Magnetic resonance studies on various nuclei in barlowite evidence that the Cu moment arrangement is far from perfect and that the magnetic excitations are unconventional in nature.

\section*{Methods}
Polycrystalline samples of Cu$_4$(OH)$_6$FBr were synthesised by hydrothermal reaction. Copper carbonate hydroxide CuCO$_3$.Cu(OH)$_2$, ammonium fluoride NH$_4$F, and  Hydrobromic acid HBr were used as starting materials and sealed in a Teflon-lined stainless steel vessel. The reaction mixture was heated at 393~K for 4 days and then cooled down to room temperature. Blue-green powder of Cu$_4$(OH)$_6$FBr were obtained after filtration and no foreign phases were observed in  x-ray diffraction.  .

Polycrystalline barlowite  samples of a few 100 mg are studied through $^1$H-, $^{19}$F-, and $^{79,81}$Br- NMR measurements. NMR measurements were performed using a home-built Techmag Apollo spectrometer with a standard local probe and a sweepable superconducting magnet. Field sweep NMR spectra were obtained by the integration of spin echo signals at a fixed frequency. Br-NQR spectra were measured using a frequency step point-by-point spin-echo technique. At each frequency point, the area under the spin-echo real part was integrated after proper phase adjustment in the time domain. Nuclear spin-lattice relaxation rates were measured using a standard inversion recovery method, where the nuclear magnetization $M(t)$ were obtained from the recovery of the spin-echo magnitude as a function of the time interval $\tau$ between the inversion pulse ($\pi$ pulse) and the $\pi/2 -\pi$ spin-echo sequence.

Density-functional calculations were performed in the full-potential local-orbital FPLO code\cite{Koepernik1999} using the Perdew-Wang flavor of the exchange-correlation potential.\cite{Perdew1992} Electric field gradient on the Br site was computed using the internal procedure of FPLO. First, calculations were performed for the experimental atomic positions\cite{Elliott2014}  with Cu2 atoms placed into one of the three symmetry-related sites around the three-fold axis. This reduced the symmetry to monoclinic $C2/m$. Additionally, full structural relaxation was performed, resulting in two minima with nearly equal energies (energy difference of less than 1~meV/f.u.) and different mutual arrangement of the Cu2 atoms. The relaxation introduced deviations from the three-fold symmetry not only at the Cu2 sites, but for all atoms throughout the structure leading to the overall monoclinic symmetry.

\section*{Acknowledgements}
The authors are grateful to H. Yasuoka, P. Mendels, and R. Stern for fruitful discussions.
\newpage
\section*{Supplementary informations}
\section*{Magnetization}
\begin{figure}[ht]
{\centering {\includegraphics[width=0.8\textwidth]{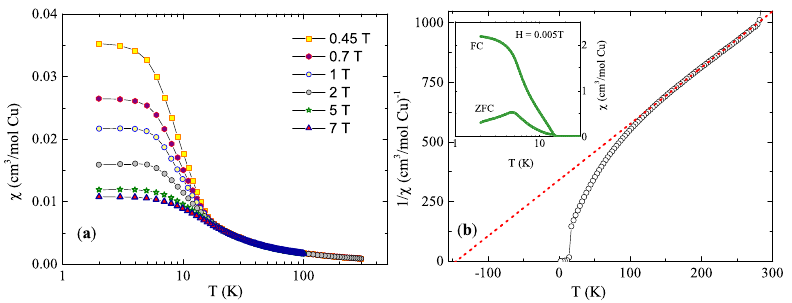}}\par} \caption{(Color online) (a) The temperature dependence of magnetic susceptibility ($\chi$) measured at different applied field. (b) Inverse magnetic susceptibility 1/$\chi$ at 0.1~T and the solid line is the Curie-Weiss fit. The inset shows the field-cooled and zero-field-cooled susceptibility measured at 0.05~T} \label{Chi}
\end{figure}

The temperature dependence of  (1.8~K$\leq T \leq$ 300~K) magnetic susceptibility ($\chi$) measured  at different applied magnetic fields ($H$) is shown in Fig.~\ref{Chi}(a). At high temperatures, $\chi(T)$ behaves in a Curie-Weiss manner and can be fitted by the expression
$$\chi=\chi_0+\frac{C}{T-\theta_{\rm CW}}$$
where $\chi_0$ is the temperature-independent contribution consists of diamagnetism of the core electron shells ($\chi_{\rm core}$) and Van-Vleck paramagnetism ($\chi_{\rm VV}$) of the open shells of the Cu$^{2+}$ ions present in the sample. The second term is the Curie-Weiss (CW) law with the Curie-Weiss temperature ($\theta_{\rm CW}$) and curie constant $C=N_{\rm A}\mu_{\rm eff}^2/3k_{\rm B}$, where $N_{\rm A}$ is Avogadro's number, $k_{\rm B}$ is the Boltzmann constant, and $\mu_{\rm eff}$ is the effective moment. The expression for effective moment is given by $\mu_{\rm eff}=g\sqrt{S(S+1)} \mu_{\rm B}$ where $g$ is the Land$\acute{e}$ g-factor and $S$ is the spin quantum number. Our fit to the 0.1~T data [Fig.~\ref{Chi}(b)] in the high temperature regime yields $\chi_0\simeq$ -1.19$\times$10$^{-6}$ cm$^3$/mol, $C\simeq$0.45~cm$^3$K/mol, and $\theta_{\rm CW}\simeq$ -145~K. The effective moment was calculated to be $\mu_{\rm eff}\simeq$ 1.9~$\mu_{\rm B}$/Cu. These values are in good agreement with the reported values.\cite{Han2014}

\begin{figure}[ht]
{\centering {\includegraphics[width=0.65\textwidth]{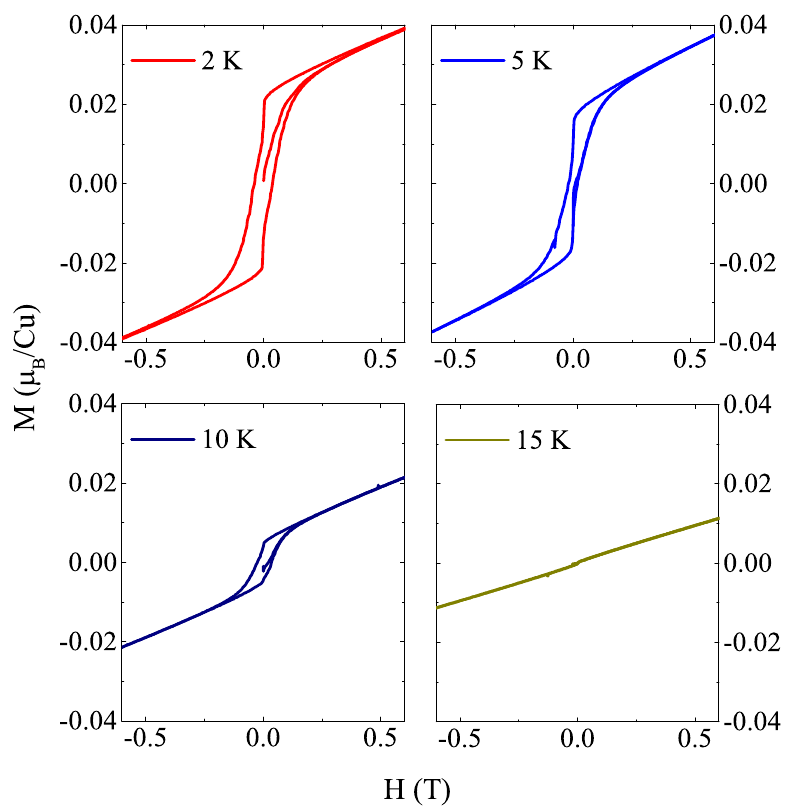}}\par} \caption{(Color online) Magnetization isotherms ($M$ vs. $H$) of measured at different temperatures.} \label{mh}
\end{figure}

 At low temperatures (1.8~K$\leq T \leq$ 30~K) zero-field-cooled (ZFC) and field-cooled (FC) magnetic susceptibility was measured at an applied field of 0.005~T. As Shown in the inset of Fig.~\ref{Chi}(b), the ZFC-FC data shows a clear splitting at 15~K which indicates a phase transition to a long-range ordered state with a small ferromagnetic moment. Isothermal magnetization $M(H)$ measured at different temperatures are shown inf Fig.~\ref{mh}. It shows a hysteresis below $T_{\rm N}$ due to the interkagome Cu or Dzyaloshinskii-Moriya interaction (DMI).

\begin{thebibliography}{10}
\expandafter\ifx\csname url\endcsname\relax
  \def\url#1{\texttt{#1}}\fi
\expandafter\ifx\csname urlprefix\endcsname\relax\def\urlprefix{URL }\fi
\expandafter\ifx\csname doiprefix\endcsname\relax\def\doiprefix{DOI }\fi
\providecommand{\bibinfo}[2]{#2}
\providecommand{\eprint}[2][]{\url{#2}}

\bibitem{Balents2010}
\bibinfo{author}{Balents, L.}
\newblock \bibinfo{journal}{\bibinfo{title}{Spin liquids in frustrated
  magnets}}.
\newblock {\emph{\JournalTitle{Nature}}} \textbf{\bibinfo{volume}{464}},
  \bibinfo{pages}{199--208} (\bibinfo{year}{2010}).

\bibitem{Savary2017}
\bibinfo{author}{Savary, L.} \& \bibinfo{author}{Balents, L.}
\newblock \bibinfo{journal}{\bibinfo{title}{Quantum spin liquids: a review}}.
\newblock {\emph{\JournalTitle{Reports on Progress in Physics}}}
  \textbf{\bibinfo{volume}{80}}, \bibinfo{pages}{016502}
  (\bibinfo{year}{2017}).

\bibitem{Zhou2017}
\bibinfo{author}{Zhou, Y.}, \bibinfo{author}{Kanoda, K.} \&
  \bibinfo{author}{Ng, T.-K.}
\newblock \bibinfo{journal}{\bibinfo{title}{Quantum spin liquid states}}.
\newblock {\emph{\JournalTitle{Rev. Mod. Phys.}}}
  \textbf{\bibinfo{volume}{89}}, \bibinfo{pages}{025003}
  (\bibinfo{year}{2017}).

\bibitem{Powell2011}
\bibinfo{author}{Powell, B.~J.} \& \bibinfo{author}{McKenzie, R.~H.}
\newblock \bibinfo{journal}{\bibinfo{title}{Quantum frustration in organic mott
  insulators: from spin liquids to unconventional superconductors}}.
\newblock {\emph{\JournalTitle{Reports on Progress in Physics}}}
  \textbf{\bibinfo{volume}{74}}, \bibinfo{pages}{056501}
  (\bibinfo{year}{2011}).

\bibitem{Helton2007}
\bibinfo{author}{Helton, J.~S.} \emph{et~al.}
\newblock \bibinfo{journal}{\bibinfo{title}{Spin dynamics of the spin-$1/2$
  kagome lattice antiferromagnet $\text{ZnCu$_{3}$(OH)$_{6}$Cl$_{2}$}$}}.
\newblock {\emph{\JournalTitle{Phys. Rev. Lett.}}}
  \textbf{\bibinfo{volume}{98}}, \bibinfo{pages}{107204}
  (\bibinfo{year}{2007}).

\bibitem{Han2012}
\bibinfo{author}{Han, T.-H.} \emph{et~al.}
\newblock \bibinfo{journal}{\bibinfo{title}{Fractionalized excitations in the
  spin-liquid state of a kagome-lattice antiferromagnet}}.
\newblock {\emph{\JournalTitle{Nature}}} \textbf{\bibinfo{volume}{492}},
  \bibinfo{pages}{406--410} (\bibinfo{year}{2012}).

\bibitem{Norman2016}
\bibinfo{author}{Norman, M.~R.}
\newblock \bibinfo{journal}{\bibinfo{title}{Colloquium:herbertsmithite and the
  search for the quantum spin liquid}}.
\newblock {\emph{\JournalTitle{Rev. Mod. Phys.}}}
  \textbf{\bibinfo{volume}{88}}, \bibinfo{pages}{041002}
  (\bibinfo{year}{2016}).

\bibitem{Yamashita2008}
\bibinfo{author}{Yamashita, S.} \emph{et~al.}
\newblock \bibinfo{journal}{\bibinfo{title}{Thermodynamic properties of a
  spin-1/2 spin-liquid state in a $\kappa$-type organic salt}}.
\newblock {\emph{\JournalTitle{Nat Phys}}} \textbf{\bibinfo{volume}{4}},
  \bibinfo{pages}{459--462} (\bibinfo{year}{2008}).

\bibitem{Yamashita2009}
\bibinfo{author}{Yamashita, M.} \emph{et~al.}
\newblock \bibinfo{journal}{\bibinfo{title}{Thermal-transport measurements in a
  quantum spin-liquid state of the frustrated triangular magnet
  -$\text{(BEDT-TTF)$_2$Cu$_2$(CN)$_3$}$}}.
\newblock {\emph{\JournalTitle{Nat Phys}}} \textbf{\bibinfo{volume}{5}},
  \bibinfo{pages}{44--47} (\bibinfo{year}{2009}).

\bibitem{Singh2013}
\bibinfo{author}{Singh, Y.}, \bibinfo{author}{Tokiwa, Y.},
  \bibinfo{author}{Dong, J.} \& \bibinfo{author}{Gegenwart, P.}
\newblock \bibinfo{journal}{\bibinfo{title}{Spin liquid close to a quantum
  critical point in $\text{Na${}_{4}$Ir${}_{3}$O${}_{8}$}$}}.
\newblock {\emph{\JournalTitle{Phys. Rev. B}}} \textbf{\bibinfo{volume}{88}},
  \bibinfo{pages}{220413} (\bibinfo{year}{2013}).

\bibitem{Han2014}
\bibinfo{author}{Han, T.-H.}, \bibinfo{author}{Singleton, J.} \&
  \bibinfo{author}{Schlueter, J.~A.}
\newblock \bibinfo{journal}{\bibinfo{title}{Barlowite: A spin-$1/2$
  antiferromagnet with a geometrically perfect kagome motif}}.
\newblock {\emph{\JournalTitle{Phys. Rev. Lett.}}}
  \textbf{\bibinfo{volume}{113}}, \bibinfo{pages}{227203}
  (\bibinfo{year}{2014}).

\bibitem{Jeschke2015}
\bibinfo{author}{Jeschke, H.~O.} \emph{et~al.}
\newblock \bibinfo{journal}{\bibinfo{title}{Barlowite as a canted
  antiferromagnet: Theory and experiment}}.
\newblock {\emph{\JournalTitle{Phys. Rev. B}}} \textbf{\bibinfo{volume}{92}},
  \bibinfo{pages}{094417} (\bibinfo{year}{2015}).

\bibitem{Liu2015}
\bibinfo{author}{Liu, Z.}, \bibinfo{author}{Zou, X.}, \bibinfo{author}{Mei,
  J.-W.} \& \bibinfo{author}{Liu, F.}
\newblock \bibinfo{journal}{\bibinfo{title}{Selectively doping barlowite for
  quantum spin liquid: A first-principles study}}.
\newblock {\emph{\JournalTitle{Phys. Rev. B}}} \textbf{\bibinfo{volume}{92}},
  \bibinfo{pages}{220102} (\bibinfo{year}{2015}).

\bibitem{Gati2015}
\bibinfo{author}{Gati, E.}, \bibinfo{author}{Wolf, B.},
  \bibinfo{author}{Schlueter, J.~A.} \& \bibinfo{author}{Lang, M.}
\newblock \bibinfo{journal}{\bibinfo{title}{Dilatometric studies on single
  crystalline barlowite – a structurally perfect spin-1/2 kagome system}}.
\newblock {\emph{\JournalTitle{Physics Procedia}}}
  \textbf{\bibinfo{volume}{75}}, \bibinfo{pages}{597 -- 604}
  (\bibinfo{year}{2015}).
\newblock \bibinfo{note}{20th International Conference on Magnetism, ICM 2015}.

\bibitem{Han2016}
\bibinfo{author}{Han, T.-H.}, \bibinfo{author}{Isaacs, E.~D.},
  \bibinfo{author}{Schlueter, J.~A.} \& \bibinfo{author}{Singleton, J.}
\newblock \bibinfo{journal}{\bibinfo{title}{Anisotropy: Spin order and
  magnetization of single-crystalline $\text{Cu$_{4}$(OH)$_{6}$FBr}$
  barlowite}}.
\newblock {\emph{\JournalTitle{Phys. Rev. B}}} \textbf{\bibinfo{volume}{93}},
  \bibinfo{pages}{214416} (\bibinfo{year}{2016}).

\bibitem{Guterding2016}
\bibinfo{author}{Guterding, D.}, \bibinfo{author}{Valent\'{\i}, R.} \&
  \bibinfo{author}{Jeschke, H.~O.}
\newblock \bibinfo{journal}{\bibinfo{title}{Reduction of magnetic interlayer
  coupling in barlowite through isoelectronic substitution}}.
\newblock {\emph{\JournalTitle{Phys. Rev. B}}} \textbf{\bibinfo{volume}{94}},
  \bibinfo{pages}{125136} (\bibinfo{year}{2016}).

\bibitem{Feng2017}
\bibinfo{author}{Feng, Z.} \emph{et~al.}
\newblock \bibinfo{journal}{\bibinfo{title}{Gapped spin-1/2 spinon excitations
  in a new kagome quantum spin liquid compound $\text{Cu$_3$Zn(OH)$_6$FBr}$}}.
\newblock {\emph{\JournalTitle{Chinese Physics Letters}}}
  \textbf{\bibinfo{volume}{34}}, \bibinfo{pages}{077502}
  (\bibinfo{year}{2017}).

\bibitem{Yuan2018}
\bibinfo{author}{Wei, Y.} \emph{et~al.}
\newblock \bibinfo{journal}{\bibinfo{title}{Evidence for a \text{Z}$_2$2
  topological ordered quantum spin liquid in a kagome-lattice
  antiferromagnet}}.
\newblock {\emph{\JournalTitle{arXiv:1710.02991v1 [cond-mat.str-el]}}} .

\bibitem{mendels2016}
\bibinfo{author}{Mendels, P.} \& \bibinfo{author}{Bert, F.}
\newblock \bibinfo{journal}{\bibinfo{title}{Quantum kagome frustrated
  antiferromagnets: One route to quantum spin liquids}}.
\newblock {\emph{\JournalTitle{Comptes Rendus Physique}}}
  \textbf{\bibinfo{volume}{17}}, \bibinfo{pages}{455--470}
  (\bibinfo{year}{2016}).

\bibitem{bert2009}
\bibinfo{author}{Bert, F.} \emph{et~al.}
\newblock \bibinfo{title}{Frustrated magnetism in the quantum kagome
  herbertsmithite $\text{ZnCu$_3$(OH)$_6$Cl$_2$}$ antiferromagnet}.
\newblock In \emph{\bibinfo{booktitle}{Journal of Physics: Conference Series}},
  vol. \bibinfo{volume}{145}, \bibinfo{pages}{012004}
  (\bibinfo{organization}{IOP Publishing}, \bibinfo{year}{2009}).

\bibitem{lacroix2013}
\bibinfo{author}{Lacroix, C.}, \bibinfo{author}{Mendels, P.} \&
  \bibinfo{author}{Mila, F.}
\newblock \emph{\bibinfo{title}{Introduction to Frustrated Magnetism:
  Materials, Experiments, Theory: Chapter-9}} (\bibinfo{publisher}{Springer},
  \bibinfo{year}{2013}).

\bibitem{Pregelj2013}
\bibinfo{author}{Pregelj, M.} \emph{et~al.}
\newblock \bibinfo{journal}{\bibinfo{title}{Evolution of magnetic and crystal
  structures in the multiferroic $\text{FeTe${}_{2}$O${}_{5}$Br}$}}.
\newblock {\emph{\JournalTitle{Phys. Rev. B}}} \textbf{\bibinfo{volume}{87}},
  \bibinfo{pages}{144408} (\bibinfo{year}{2013}).

\bibitem{Comment2010}
\bibinfo{author}{Comment, A.} \emph{et~al.}
\newblock \bibinfo{journal}{\bibinfo{title}{$\text{NMR and NQR}$ study of the
  tetrahedral frustrated quantum spin system
  $\text{Cu$_{2}$Te$_{2}$O$_{5}$Br$_{2}$}$ in its paramagnetic phase}}.
\newblock {\emph{\JournalTitle{Phys. Rev. B}}} \textbf{\bibinfo{volume}{82}},
  \bibinfo{pages}{214416} (\bibinfo{year}{2010}).

\bibitem{Baek2012}
\bibinfo{author}{Baek, S.-H.}, \bibinfo{author}{Choi, K.-Y.},
  \bibinfo{author}{Berger, H.}, \bibinfo{author}{B\"uchner, B.} \&
  \bibinfo{author}{Grafe, H.-J.}
\newblock \bibinfo{journal}{\bibinfo{title}{Persistence of singlet fluctuations
  in the coupled spin tetrahedra system
  $\text{Cu${}_{2}$Te${}_{2}$O${}_{5}$Br${}_{2}$}$ revealed by high-field
  magnetization, ${}^{79}$$\text{Br NQR, and ${}^{125}$Te NMR}$}}.
\newblock {\emph{\JournalTitle{Phys. Rev. B}}} \textbf{\bibinfo{volume}{86}},
  \bibinfo{pages}{180405} (\bibinfo{year}{2012}).

\bibitem{Rebecca2009}
\bibinfo{author}{Chapman, R.~P.}, \bibinfo{author}{Widdifield, C.~M.} \&
  \bibinfo{author}{Bryce, D.~L.}
\newblock \bibinfo{journal}{\bibinfo{title}{Solid-state nmr of quadrupolar
  halogen nuclei}}.
\newblock {\emph{\JournalTitle{Progress in Nuclear Magnetic Resonance
  Spectroscopy}}} \textbf{\bibinfo{volume}{55}}, \bibinfo{pages}{215 -- 237}
  (\bibinfo{year}{2009}).

\bibitem{Procissi2004}
\bibinfo{author}{Procissi, D.} \emph{et~al.}
\newblock \bibinfo{journal}{\bibinfo{title}{Magnetic susceptibility and spin
  dynamics of a polyoxovanadate cluster: A proton nmr study of a model spin
  tetramer}}.
\newblock {\emph{\JournalTitle{Phys. Rev. B}}} \textbf{\bibinfo{volume}{69}},
  \bibinfo{pages}{094436} (\bibinfo{year}{2004}).

\bibitem{Suh2003}
\bibinfo{author}{Suh, B.~J.} \emph{et~al.}
\newblock \bibinfo{journal}{\bibinfo{title}{Magnetic properties and spin
  dynamics in magnetic molecule {Mn3}}}.
\newblock {\emph{\JournalTitle{J. Appl. Phys.}}} \textbf{\bibinfo{volume}{93}},
  \bibinfo{pages}{7098--7100} (\bibinfo{year}{2003}).

\bibitem{Ranjith2015}
\bibinfo{author}{Ranjith, K.~M.}, \bibinfo{author}{Majumder, M.},
  \bibinfo{author}{Baenitz, M.}, \bibinfo{author}{Tsirlin, A.~A.} \&
  \bibinfo{author}{Nath, R.}
\newblock \bibinfo{journal}{\bibinfo{title}{Frustrated three-dimensional
  antiferromagnet $\text{Li$_2$CuW$_2$O$_8$}$: $^{7}$$\text{Li}$ nmr and the
  effect of nonmagnetic dilution}}.
\newblock {\emph{\JournalTitle{Phys. Rev. B}}} \textbf{\bibinfo{volume}{92}},
  \bibinfo{pages}{024422} (\bibinfo{year}{2015}).

\bibitem{Belesi2007}
\bibinfo{author}{Belesi, M.}, \bibinfo{author}{Zong, X.},
  \bibinfo{author}{Borsa, F.}, \bibinfo{author}{Milios, C.~J.} \&
  \bibinfo{author}{Perlepes, S.~P.}
\newblock \bibinfo{journal}{\bibinfo{title}{Proton nmr study in hexanuclear
  manganese single-molecule magnets}}.
\newblock {\emph{\JournalTitle{Phys. Rev. B}}} \textbf{\bibinfo{volume}{75}},
  \bibinfo{pages}{064414} (\bibinfo{year}{2007}).

\bibitem{Slichter1990}
\bibinfo{author}{Slichter, C.~P.}
\newblock \emph{\bibinfo{title}{Principles of Magnetic Resonance}}
  (\bibinfo{publisher}{Springer}, \bibinfo{address}{Berlin},
  \bibinfo{year}{1990}).

\bibitem{Abragam1996}
\bibinfo{author}{Abragam, A.}
\newblock \emph{\bibinfo{title}{Principles of Nuclear Magnetism}}
  (\bibinfo{publisher}{Oxford Univesity press}, \bibinfo{address}{Oxford},
  \bibinfo{year}{1996}).

\bibitem{Grafe2008}
\bibinfo{author}{Grafe, H.-J.} \emph{et~al.}
\newblock \bibinfo{journal}{\bibinfo{title}{$^{75}\mathrm{As}$ nmr studies of
  superconducting $\text{LaFeAsO$_{0.9}$F$_{0.1}$}$}}.
\newblock {\emph{\JournalTitle{Phys. Rev. Lett.}}}
  \textbf{\bibinfo{volume}{101}}, \bibinfo{pages}{047003}
  (\bibinfo{year}{2008}).

\bibitem{Das1958}
\bibinfo{author}{Das, T.~P.} \& \bibinfo{author}{Hahn, E.~L.}
\newblock \emph{\bibinfo{title}{Solid State Physics}}.
\newblock Suppl. 1 (\bibinfo{publisher}{Academic Press}, \bibinfo{address}{New
  York}, \bibinfo{year}{1958}).

\bibitem{Pasco2018}
\bibinfo{author}{Pasco, C.~M.} \emph{et~al.}
\newblock \bibinfo{journal}{\bibinfo{title}{Single crystal growth of
  $\text{Cu$_{4}$(OH)$_{6}$FBr}$ and universal behavior in quantum spin liquid
  candidates synthetic barlowite and herbertsmithite}}.
\newblock {\emph{\JournalTitle{arXiv:1801.05769 [cond-mat.str-el]}}} .

\bibitem{Feng2018}
\bibinfo{author}{Feng, Z.} \emph{et~al.}
\newblock \bibinfo{journal}{\bibinfo{title}{Effect of \text{Zn} doping on the
  antiferromagnetism in kagome $\text{Cu$_{4-x}$Zn$_{x}$(OH)$_{6}$FBr}$}}.
\newblock {\emph{\JournalTitle{arXiv:1712.06732v2 [cond-mat.str-el]}}} .

\bibitem{Moriya1956}
\bibinfo{author}{Moriya, T.}
\newblock \bibinfo{journal}{\bibinfo{title}{Nuclear magnetic relaxation in
  antiferromagnetics}}.
\newblock {\emph{\JournalTitle{Progress of Theoretical Physics}}}
  \textbf{\bibinfo{volume}{16}}, \bibinfo{pages}{23--44}
  (\bibinfo{year}{1956}).

\bibitem{Moriya1985}
\bibinfo{author}{Moriya, T.}
\newblock \emph{\bibinfo{title}{Spin Fluctuations in Itinerant Electron
  Magnetism}} (\bibinfo{publisher}{Springer}, \bibinfo{address}{Berlin},
  \bibinfo{year}{1985}).

\bibitem{Imai2008}
\bibinfo{author}{Imai, T.}, \bibinfo{author}{Nytko, E.~A.},
  \bibinfo{author}{Bartlett, B.~M.}, \bibinfo{author}{Shores, M.~P.} \&
  \bibinfo{author}{Nocera, D.~G.}
\newblock \bibinfo{journal}{\bibinfo{title}{$^{63}\mathrm{Cu}$,
  $^{35}\mathrm{Cl}$, and $^{1}\mathrm{H}$ nmr in the $\mathrm{S} =
  \frac{1}{2}$ kagome lattice $\text{ZnCu$_3$(OH)$_6$Cl$_2$}$}}.
\newblock {\emph{\JournalTitle{Phys. Rev. Lett.}}}
  \textbf{\bibinfo{volume}{100}}, \bibinfo{pages}{077203}
  (\bibinfo{year}{2008}).

\bibitem{Itou2010}
\bibinfo{author}{Itou, T.}, \bibinfo{author}{Oyamada, A.},
  \bibinfo{author}{Maegawa, S.} \& \bibinfo{author}{Kato, R.}
\newblock \bibinfo{journal}{\bibinfo{title}{Instability of a quantum spin
  liquid in an organic triangular-lattice antiferromagnet}}.
\newblock {\emph{\JournalTitle{Nature Physics}}} \textbf{\bibinfo{volume}{6}},
  \bibinfo{pages}{673} (\bibinfo{year}{2010}).

\bibitem{Shiroka2011}
\bibinfo{author}{Shiroka, T.} \emph{et~al.}
\newblock \bibinfo{journal}{\bibinfo{title}{Distribution of nmr relaxations in
  a random heisenberg chain}}.
\newblock {\emph{\JournalTitle{Phys. Rev. Lett.}}}
  \textbf{\bibinfo{volume}{106}}, \bibinfo{pages}{137202}
  (\bibinfo{year}{2011}).

\bibitem{Bosio2017}
\bibinfo{author}{Bosio\ifmmode \check{c}\else
  \v{c}\fi{}i\ifmmode~\acute{c}\else \'{c}\fi{}, M.} \emph{et~al.}
\newblock \bibinfo{journal}{\bibinfo{title}{Possible quadrupolar nematic phase
  in the frustrated spin chain $\text{LiCuSbO$_{4}$}$: An nmr investigation}}.
\newblock {\emph{\JournalTitle{Phys. Rev. B}}} \textbf{\bibinfo{volume}{96}},
  \bibinfo{pages}{224424} (\bibinfo{year}{2017}).

\bibitem{Shockley2015}
\bibinfo{author}{Shockley, A.~C.}, \bibinfo{author}{Bert, F.},
  \bibinfo{author}{Orain, J.-C.}, \bibinfo{author}{Okamoto, Y.} \&
  \bibinfo{author}{Mendels, P.}
\newblock \bibinfo{journal}{\bibinfo{title}{Frozen state and spin liquid
  physics in $\text{Na$_{4}$Ir$_{3}$O$_{8}$}$: An nmr study}}.
\newblock {\emph{\JournalTitle{Phys. Rev. Lett.}}}
  \textbf{\bibinfo{volume}{115}}, \bibinfo{pages}{047201}
  (\bibinfo{year}{2015}).

\bibitem{Khuntia2017}
\bibinfo{author}{Khuntia, P.} \emph{et~al.}
\newblock \bibinfo{journal}{\bibinfo{title}{Local magnetism and spin dynamics
  of the frustrated honeycomb rhodate $\text{Li$_{2}$RhO$_{3}$}$}}.
\newblock {\emph{\JournalTitle{Phys. Rev. B}}} \textbf{\bibinfo{volume}{96}},
  \bibinfo{pages}{094432} (\bibinfo{year}{2017}).

\bibitem{Moriya1963}
\bibinfo{author}{Moriya, T.}
\newblock \bibinfo{journal}{\bibinfo{title}{The effect of electron-electron
  interaction on the nuclear spin relaxation in metals}}.
\newblock {\emph{\JournalTitle{Journal of the Physical Society of Japan}}}
  \textbf{\bibinfo{volume}{18}}, \bibinfo{pages}{516--520}
  (\bibinfo{year}{1963}).

\bibitem{Koepernik1999}
\bibinfo{author}{Koepernik, K.} \& \bibinfo{author}{Eschrig, H.}
\newblock \bibinfo{journal}{\bibinfo{title}{Full-potential nonorthogonal
  local-orbital minimum-basis band-structure scheme}}.
\newblock {\emph{\JournalTitle{Phys. Rev. B}}} \textbf{\bibinfo{volume}{59}},
  \bibinfo{pages}{1743--1757} (\bibinfo{year}{1999}).

\bibitem{Perdew1992}
\bibinfo{author}{Perdew, J.~P.} \& \bibinfo{author}{Wang, Y.}
\newblock \bibinfo{journal}{\bibinfo{title}{Accurate and simple analytic
  representation of the electron-gas correlation energy}}.
\newblock {\emph{\JournalTitle{Phys. Rev. B}}} \textbf{\bibinfo{volume}{45}},
  \bibinfo{pages}{13244--13249} (\bibinfo{year}{1992}).

\bibitem{Elliott2014}
\bibinfo{author}{Elliott, P.}, \bibinfo{author}{Cooper, M.~A.} \&
  \bibinfo{author}{Pring, A.}
\newblock \bibinfo{journal}{\bibinfo{title}{Barlowite,
  $\text{Cu$_{4}$(OH)$_{6}$FBr}$, a new mineral isotructural with
  claringbullite: description and crystal structure}}.
\newblock {\emph{\JournalTitle{Mineralogical Magazine}}}
  \textbf{\bibinfo{volume}{78}}, \bibinfo{pages}{1755} (\bibinfo{year}{2014}).

\end{thebibliography}
\end{document}